\documentclass[prl,nofootinbib,twocolumn]{revtex4}
\def\mysection#1{{\bf #1.} }
\def\mysections#1{{\bf #1.} }

\usepackage{amssymb}
\usepackage{amsmath}
\usepackage[dvips]{graphicx}
\usepackage{longtable}
\usepackage{verbatim}
\usepackage{amsfonts}

\newcommand{\be}{\begin{equation}}
\newcommand{\ee}{\end{equation}}
\newcommand{\bea}{\begin{eqnarray}}
\newcommand{\eea}{\end{eqnarray}}
\newcommand{\beq}{\begin{equation}}
\newcommand{\eeq}{\end{equation}}
\def\beqa{\begin{eqnarray}}
\def\eeqa{\end{eqnarray}}
\newcommand{\no}{\nonumber}
\def\lsim{\mathrel{\rlap{\lower4pt\hbox{\hskip1pt$\sim$}}
    \raise1pt\hbox{$<$}}}         
\def\gsim{\mathrel{\rlap{\lower4pt\hbox{\hskip1pt$\sim$}}
    \raise1pt\hbox{$>$}}}         

\begin{document}

\vspace*{-30mm}

\title{\boldmath Probing the Seesaw and Gauge Mediation Scales with
 BR($\mu\to e\gamma$) and $|U_{e3}|$}

\author{Daniel Grossman}\email{daniel.grossman@weizmann.ac.il}
\affiliation{Department of Particle Physics and Astrophysics,
  Weizmann Institute of Science, Rehovot 76100, Israel}

\author{Yosef Nir}\email{yosef.nir@weizmann.ac.il}
\affiliation{Department of Particle Physics and Astrophysics,
  Weizmann Institute of Science, Rehovot 76100, Israel}

\vspace*{1cm}

\begin{abstract}
  The new MEG bound on BR($\mu\to e\gamma$) provides the strongest
  upper bound on the scale of gauge mediation of supersymmetry
  breaking. If, in the future, this decay is observed by MEG, the
  mediation scale will become known to within one order of magnitude,
  and the seesaw scale will be constrained. In such a case,
  contributions from Planck mediated supersymmetry breaking are likely
  to be non-negligible, and an interpretation in terms of purely
  seesaw parameters will be impossible. The recent evidence for
  $|U_{e3}|\sim0.15$ further sharpens the predictions of gauge mediated
  supersymmetry breaking.
\end{abstract}

\maketitle

\mysection{Introduction}
The MEG experiment has recently announced a new bound
\cite{Adam:2011ch},
\beq\label{eq:mega}
{\rm BR}(\mu\to e\gamma)\leq2.4\times10^{-12},
\eeq
a factor of 5 improvement over the previous MEGA bound
\cite{Brooks:1999pu}. Within a few years, MEG is expected to explore
the range of
\beq\label{eq:meg}
{\rm BR}(\mu\to e\gamma)\gsim10^{-13}.
\eeq
The Standard Model, extended with the seesaw mechanism to account for
neutrino masses, predicts a branching ratio that is about forty orders
of magnitude lower. Thus, future observation of BR($\mu\to e\gamma$)
within the range of $2.4\times10^{-12}-10^{-13}$ will signal new
physics. A leading candidate to account for such a rate will be the
supersymmetric standard model, where lepton flavor violation arises,
in general, in the soft supersymmetry breaking slepton mass-squared
terms. The implications will be particularly interesting in the
framework of models with universal soft terms, such as gauge mediated
supersymmetry breaking (GMSB), where the supersymmetric flavor problem
is naturally solved. In such a framework, lepton flavor violation can
still arise via the renormalization group evolution (RGE) effect of
the singlet neutrinos on the soft breaking terms
\cite{NYU/TR4/86,Tobe:2003nx}. (For reviews, see for example
\cite{hep-ph/0407325,arXiv:0801.1826}.)

An observation of $\mu\to e\gamma$, when interpreted within the GMSB
framework, will put a lower bound on the seesaw scale. The lower the
seesaw scale, the smaller the neutrino Yukawa couplings, leading
eventually to negligible RGE effects and to a decay rate below the
experimental sensitivity. A GMSB interpretation of such an observation
will also put an upper bound on the seesaw scale. It must be lower
than the mediation scale in order for the RGE effects to take place.
The mediation scale itself is bounded from above, or else Planck scale
mediation (PMSB), where the flavor structure is not expected to be
universal, becomes significant. Conversely, an upper bound on the rate
of $\mu\to e\gamma$ provides an upper bound on the scale of gauge
mediation, to avoid large contributions from PMSB, and an upper bound
on the seesaw scale, if it is lower than the mediation scale.

The purpose of this work is to obtain the constraints on the seesaw
and mediation scales that follow from the new bound (\ref{eq:mega}).
We further study the constraints that will follow from observing
$\mu\to e\gamma$, to understand whether such an observation can be
translated into constraints on the seesaw flavor parameters, and to
explore further phenomenological implications.

\mysection{Low energy neutrino parameters}
The GMSB predictions for lepton flavor violation depend also on the
low energy neutrino parameters. Specifically, the dependence is on the
three following combinations:
\beqa\label{eq:muij}
\mu_{\mu e}&=&s_{12}c_{12}c_{23}(m_2-m_1)+s_{13}s_{23}(m_3-m_1),\no\\
\mu_{\tau e}&=&-s_{12}c_{12}s_{23}(m_2-m_1)+s_{13}c_{23}(m_3-m_1),\no\\
\mu_{\tau\mu}&=&c_{23}s_{23}[-c_{12}^2(m_2-m_1)+(m_3-m_1)],
\eeqa
where $s_{ij}\equiv\sin\theta_{ij}$, $c_{ij}\equiv\cos\theta_{ij}$,
with $\theta_{ij}$ the angles of the leptonic mixing matrix $U$, and
we use $c_{13}=1$. To evaluate the numerical values of the $\mu_{ij}$
parameters, we assume normal hierarchy for the neutrino masses,
whereby $m_2-m_1\simeq\sqrt{\Delta m^2_{21}}$ and
$m_3-m_1\simeq\sqrt{\Delta m^2_{31}}$ \cite{pdg}:
\beqa\label{eq:nummu}
m_2-m_1=0.009\ {\rm eV},\ \ \
  m_3-m_1=0.05\ {\rm eV}.
\eeqa
The recent intriguing measurements of $|U_{e3}|$ by the T2K
\cite{Abe:2011sj} and MINOS \cite{Adamson:2011qu} experiments have been
combined into a global analysis of oscillation data
\cite{Fogli:2011qn}, yielding
\beq\label{eq:t2k}
s_{13}\simeq0.15\pm0.03\ (1\sigma).
\eeq
For the other two mixing angles, we use the central values from a
recent fit \cite{GonzalezGarcia:2010er}:
\beqa
s_{12}=0.56,\ \ \ s_{23}=0.68.
\eeqa
We obtain (taking into account only the $s_{13}$-related uncertainty):
\beqa\label{eq:muijtbm}
\mu_{\mu e}&=&0.0082\pm0.0010\ {\rm eV},\no\\
\mu_{\tau e}&=&0.0027\pm0.0011\ {\rm eV},\no\\
\mu_{\tau\mu}&=&0.022\ {\rm eV}.
\eeqa
%

\mysection{The Model}
In minimal GMSB models
\cite{hep-ph/9303230,hep-ph/9408384,hep-ph/9507378,hep-ph/9801271}
(for earlier work, see
\cite{Dine:1981gu,Nappi:1982hm,AlvarezGaume:1981wy}), 
slepton-doublet masses are given, at the messenger scale, by
\beq\label{eq:mslepton}
\hat m^2_{\widetilde L}\equiv m^2_{\widetilde
  L}(M_m)=\frac{3n_5(5\alpha_2^2+\alpha_1^2)M_s^2}
{160\pi^2},
\eeq
where $M_m$ is the mass scale of the messenger fields (that is,
the mediation scale), $M_s$ is a scale of order a (few) hundred
TeV, and $n_5$ is the number of messenger fields in the $5+\bar 5$
representation of $SU(5)$. (In this work, we focus on models with
weakly coupled messenger fields. We will explore more general models
of gauge mediation \cite{arXiv:0801.3278} in future work.)

The leptonic part of the superpotential reads
\beq\label{eq:wn}
W_\ell=E_i^c\lambda^{ij}_e L_j H_1+N_i^c\lambda^{ij}_\nu L_j H_2
-\frac12 M_N^{ij}N_i^c N_j^c,
\eeq
where $L_i$, $E_i^c$ and $N_i^c$ are leptonic superfields in the
$(2)_{-1/2}$, $(1)_{+1}$ and $(1)_0$ representations of
$SU(2)\times U(1)$, respectively, and $H_{1,2}$ are the Higgs
superfields in the $(2)_{\mp1/2}$ representations. The matrices
$\lambda_e$ and $\lambda_\nu$ are complex $3\times3$
Yukawa matrices, while $M_N$ is the symmetric Majorana mass
matrix. Without loss of generality, we work in a basis where
$\lambda_e$ and $M_N$ are diagonal.

The light neutrino Majorana mass matrix is given by
\beq\label{eq:mnu}
M_\nu=v_2^2\lambda_\nu^T M_N^{-1}\lambda_\nu,
\eeq
where $v_i=\langle H_i^0\rangle$. Diagonalization of $M_\nu$
leads to the mass eigenvalues $m_i$, $i=1,2,3$, and to the
leptonic mixing matrix $U$.

If some of the eigenvalues of $M_N$ are lower than $M_m$, then RGE
of $m^2_{\widetilde L}$ between the scales $M_m$ and $M_N$ will
generate off-diagonal terms in $m^2_{\widetilde L}$. In a generic
supersymmetric model, the RGE is given by
\cite{Hisano:1995nq,Hisano:1995cp}
\beqa\label{eq:rge}
\mu\frac{d}{d\mu}\left(m^2_{\tilde L}\right)_{ij}&=&
\left.\mu\frac{d}{d\mu}\left(m^2_{\widetilde L}\right)_{ij}
\right|_{\rm MSSM}\\
&+&\frac{1}{16\pi^2}\left[(m^2_{\widetilde L}
\lambda_\nu^\dagger\lambda_\nu
+\lambda_\nu^\dagger\lambda_\nu m^2_{\widetilde L})_{ij}\right.\no\\
&&\left.+2(\lambda_\nu^\dagger m^2_{\widetilde N}\lambda_\nu
+m^2_{\widetilde H_u}\lambda_\nu^\dagger\lambda_\nu
+A_\nu^\dagger A_\nu)_{ij}\right].\no
\eeqa
In GMSB models, the trilinear scalar couplings are negligible, $A_\nu=0$,
and the soft masses-squared of singlet fields vanish, $m_{\tilde N}=0$.
Moreover, at the messenger scale $m^2_{\tilde L}=m^2_{\tilde H_u}$.
For simplicity, we adopt from here on the {\it minimal lepton flavor
violation} (MLFV) ansatz of Ref. \cite{Cirigliano:2005ck} and take the
singlet neutrinos to be degenerate,
\beq\label{eq:mn}
M_N=m_N{\bf 1}.
\eeq
In the leading log approximation, and with the GMSB boundary conditions
(\ref{eq:mslepton}), the off-diagonal elements of the doublet slepton mass
matrix at low energy are given by
\beq\label{eq:mlga}
(m^2_{\tilde L})_{ij}=-\frac{\hat m^2_{\tilde L}}{4\pi^2}
(\lambda_\nu^\dagger\lambda_\nu)_{ij}\ln\frac{M_m}{m_N}.
\eeq

We now introduce yet another MLFV ansatz, and take the Yukawa matrix
$\lambda_\nu$ to be real \cite{Cirigliano:2005ck}. Then Eq. (\ref{eq:mnu})
leads to
\beq\label{eq:mlmnu}
(\lambda_\nu^\dagger\lambda_\nu)_{ij}=\frac{m_N \mu_{ij}}{v_2^2},
\eeq
where the $\mu_{ij}$ parameters are defined in Eq. (\ref{eq:muij}).

In calculations of the supersymmetric contributions to
flavor changing neutral current (FCNC) processes, the following
dimensionless parameters are useful:
\beq\label{eq:delij}
\hat{\delta}_{ij}\equiv\frac{(m^2_{\tilde L})_{ij}}{\hat{m}^2_{\tilde L}}
=\frac{m_N\mu_{ij}}{4\pi^2v_2^2}\ln\frac{M_m}{m_N}.
\eeq

In a full model of GMSB, one usually introduces a SM-singlet chiral
superfield $S$, whose scalar and auxiliary components acquire VEVs,
denoted by $s$ and $F_s$, respectively. The scale $M_s$ introduced in
Eq. (\ref{eq:mslepton}), which sets the size of the soft terms from
gauge mediation, and the mass scale of the messenger fields, $M_m$,
are given by
\beq
M_s\sim F_s/s,\ \ \ M_m\sim s.
\eeq
$F_s$ generates also $M_{\rm Pl}$-suppressed contributions to the soft
breaking terms. The scale of these Planck mediated supersymmetry
breaking (PMSB) contributions is given by $F_s/M_{\rm Pl}$. The PMSB
contributions have an unknown flavor structure. From here on we take
this structure to be anarchical \cite{Hiller:2008sv,Hiller:2010dv}, with
coefficients of order one:
\beq\label{eq:mgrav}
(m^2_{\widetilde L})_{ij}^{\rm grav}\sim
\left(\frac{M_s M_m}{M_{\rm Pl}}\right)^2.
\eeq
Eq. (\ref{eq:mgrav}) constitutes, on one hand, an approximate upper
bound and, on the other, a plausible estimate. Moreover, it holds also
for models where the structure of the soft breaking terms is dictated
by a Froggatt-Nielsen (FN) symmetry
\cite{Froggatt:1978nt,Nir:1993mx,Leurer:1993gy,Grossman:1995hk} with
equal FN charges for the three left-handed lepton doublets, as
suggested by the order-one leptonic mixing angles.

Generically, PMSB contributions with unknown flavor structure are
unavoidable. Consequently, constraints from flavor changing processes
provide an upper bound on $M_m$. Ref. \cite{Hiller:2010dv} analyzes
flavor violation in the quark sector and obtains
\beq\label{eq:upmm}
M_m/M_{\rm Pl}\lsim10^{-3}
\eeq
for models with anarchical PMSB terms.

The ratio between the PMSB and the GMSB
contributions to the off-diagonal elements of the doublet slepton
masses can then be estimated as
\beqa\label{eq:grga}
x_{ij}&\equiv&\frac{(m^2_{\tilde L})_{ij}^{\rm grav}}
{(m^2_{\widetilde L})_{ij}^{\rm gauge}}\sim\frac{M_m^2}{M_{\rm Pl}^2}
\frac{640\pi^4}{3n_5(5\alpha_2^2+\alpha_1^2)}\frac{1}
{(\lambda_\nu^\dagger\lambda_\nu)_{ij}\ln\frac{M_m}{m_N}}\no\\
&=&\frac{v_2^2}{M_{\rm Pl}^2}
\frac{640\pi^4}{3n_5(5\alpha_2^2+\alpha_1^2)}\frac{M_m^2}
{m_N\mu_{ij}\ln\frac{M_m}{m_N}}\\
&=&\frac{1.5}{n_5}\left(\frac{M_m}{5\times10^{13}\ {\rm GeV}}\right)^2
\frac{10^{12}\ {\rm GeV}}{m_N}
\frac{0.008\ {\rm eV}}{\mu_{ij}}
\frac{4}{\ln\frac{M_m}{m_N}}.\no
\eeqa
%

\mysection{The $\mu\to e\gamma$ decay}
Within our framework, where the main features of the spectrum are
determined by GMSB with a perturbative messenger sector, the $\mu\to
e\gamma$ decay is dominated by the chargino/sneutrino loop diagrams.
An approximate expression for the branching ratio can be written
(based on Ref. \cite{Hisano:1995cp}) as follows:
\beq\label{eq:brmeg}
{\rm BR}(\mu\to e\gamma)\simeq\frac{6\pi\alpha\alpha_2^2 t_\beta^2
  v^4}{m_{\tilde{L}}^4}h\left[\left(\frac{M_2}{m_{\tilde{L}}}\right)^2,
  \left(\frac{\mu}{m_{\tilde{L}}}\right)^2\right]\delta_{\mu e}^2,
\eeq
where $t_\beta\equiv\tan\beta$, $M_2$ is the Wino mass, $\mu$ is the
Higgsino mass term, $m_{\tilde{L}}$ is the average mass of the
quasi-degenerate sneutrinos,
\beq\label{eq:delmeg}
\delta_{\mu e}\equiv\frac{(m_{\tilde{L}}^2)_{21}}{m_{\tilde{L}}^2}
=\frac{\hat{m}_{\tilde{L}}^2}{m_{\tilde{L}}^2}\hat\delta_{\mu e},
\eeq
and
\bea\label{eq:deff}
\nonumber h(x,y)&=&\frac{xy}{(x-y)^2}
\left[g\left(x\right)-g\left(y\right)\right]^2,\\
g(x)&=&\frac{-5+4x+x^2-2\ln x-4x\ln x}{(x-1)^4}.
\eea
We note that for very large $n_5$, the approximation of Eq.
(\ref{eq:brmeg}) becomes less accurate, as the contributions from
neutralino/charged-slepton loop diagrams become non-negligible. The
bounds of Eqs. (\ref{eqn:upperMmn5eq1}) and (\ref{eq:upimp}) below are
derived from numerical calculations (based on Ref.
\cite{Petcov:2003zb}) which do take into account these additional
contributions.

An upper bound on the radiative decay,
\beq
{\rm BR}\left(\mu\rightarrow e\gamma\right)<\overline{Br},
\eeq
implies
\beq\label{eq:updel}
\delta_{\mu e}^L<6\cdot 10^{-5}\left(\frac{20}{t_{\beta}}\right)
\sqrt{\frac{0.1}{h}}\left(\frac{m_{\tilde{L}}}{300\;{\rm GeV}}\right)^2
\sqrt{\frac{\overline{Br}}{2.4\cdot 10^{-12}}}.
\eeq
Using Eqs. (\ref{eq:mgrav}) and (\ref{eq:mslepton}), we then obtain an
upper bound on the scale of gauge mediation (for anarchical PMSB
slepton mass-squared terms):
\bea\label{eq:upmmgm}
\nonumber\frac{M_m}{M_{Pl}}&<&3\cdot
10^{-5}\left(\frac{\overline{Br}}{2.4\cdot 10^{-12}}\right)^{{1}/{4}}
\left(\frac{\hat{m}_{\tilde{L}}}{300\;{\rm GeV}}\right)\\
&\times&\left(\frac{m_{\tilde{L}}}{\hat{m}_{\tilde{L}}}\right)^2
\left(\frac{20n_5}{t_{\beta}}\right)^{1/2}
\left(\frac{0.1}{h}\right)^{{1}/{4}}.
\eea
The properties of the GMSB spectrum \cite{Dimopoulos:1996yq}, and in
particular the fact that the ratio between the wino and
doublet-slepton masses increases with $n_5$, lead to \beq h\sim
0.1\to0.01\ {\rm for}\ n_5=1\to20.  \eeq The running of
$m_{\tilde{L}}^2$ is dominated by gaugino masses, such that for fixed
$\hat{m}_{\tilde{L}}$
\beq
\frac{m_{\tilde{L}}}{\hat{m}_{\tilde{L}}}\approx1.01
+\frac{9n_5}{20\pi^2}.
\eeq
Plugging these results back into Eq. (\ref{eq:upmmgm}) we obtain for
the two extreme cases of $n_5=1,20$:
\beq
\label{eqn:upperMmn5eq1}
\frac{M_m}{M_{\rm Pl}}\lesssim\begin{cases}
    3\times10^{-5}&n_5=1,\\
    1\times10^{-3}&n_5=20.\end{cases}
\eeq
These bounds scale like $\sqrt{\frac{20}{t_{\beta}}}
\left(\frac{\hat{m}_{\tilde{L}}}{300\;{\rm GeV}}\right)
\left(\frac{\overline{Br}}{2.4\cdot 10^{-12}}\right)^{\frac{1}{4}}$.
They are to be compared with Eq. (\ref{eq:upmm}). We conclude that the
new MEG bound on $\mu\to e\gamma$ provides the strongest upper bound
on the scale of gauge mediation.

Inserting Eq. (\ref{eq:delij}) into the expression (\ref{eq:brmeg}),
we obtain, for the purely GMSB contributions and for $M_m>m_N$,
\beqa\label{eq:brmegmm}
{\rm BR}(\mu\to e\gamma)&\simeq&\frac{3\alpha^3h}{2\pi^3 s^2_{2\beta}
  s_W^4}
\frac{\hat{m}_{\tilde{L}}^4m_N^2\mu_{\mu e}^2}{m_{\tilde{L}}^8}
\left(\ln\frac{M_M}{m_N}\right)^2\\
\nonumber =5&\times& 10^{-13}\left(
  \frac{\hat{m}_{\tilde{L}}}{m_{\tilde{L}}}\right)^8
\left(\frac{0.1}{s_{2\beta}}\right)^2\left(\frac{h}{0.1}\right)\\
\nonumber&\times&\left(\frac{m_N}{10^{12}\; {\rm GeV}}\right)^2
\left(\frac{\mu_{\mu e}}{0.008\; {\rm eV}}\right)^2\\
\nonumber&\times&\left(\frac{300\; {\rm GeV}}{\hat{m}_{\tilde{L}}}\right)^4
\left(\frac{\ln\frac{M_m}{m_N}}{4}\right)^2.
\eeqa
Using the expression (\ref{eq:brmegmm}), the upper bound of
Eq. (\ref{eq:mega}) implies that
\beqa\label{eq:upimp}
m_N&\gsim&M_m\ {\rm or}\\
m_N&\lsim&\begin{cases}
  2.7\times10^{12}\ {\rm GeV}&n_5=1,\\
  1.1\times10^{14}\ {\rm GeV}&n_5=20,\end{cases}\no
\eeqa
where the bounds scale like  $({\sin
    2\beta}/{0.1})({\hat{m}_{\tilde{L}}}/{300\ {\rm
    GeV}})^2$.

In the literature, the less well-motivated case of universal PMSB
(known as mSUGRA or CMSSM) is often considered
\cite{arXiv:0904.2080,Paradisi:2005fk,Arganda:2005ji}. The
supersymmetric spectrum is different from the GMSB framework, and the
approximation (\ref{eq:brmeg}) is replaced by
\beq\label{eq:brmegp}
{\rm BR}(\mu\to e\gamma)\simeq\frac{8\pi\alpha\alpha_2^2 t_\beta^2
v^4}{75\tilde m^4}\delta_{\mu e}^2,
\eeq
where $\tilde m$ is the scale of the soft breaking parameters. The
upper bound on BR($\mu\to e\gamma$) of Eq. (\ref{eq:mega}) provides an
upper bound on $\delta_{\mu e}$:
\beq\label{eq:updelp}
\delta_{\mu e}\leq1.4\times10^{-4}\ ({20}/{t_\beta})\
({\tilde m}/{300\ {\rm GeV}})^2.
\eeq
An important feature of this scenario is that the seesaw scale, $m_N$,
is necessarily below the mediation scale, $M_{\rm Pl}$. Thus, the
generation of off-diagonal entries in the doublet slepton mass-squared
matrix by RGE from $M_{\rm Pl}$ to $m_N$ is unavoidable. Consequently,
Eq. (\ref{eq:mega}) provides an upper bound on $m_N$. Using
Eq. (\ref{eq:delij}) with $M_m$ replaced by $M_{\rm Pl}$,
Eq. (\ref{eq:updelp}) yields, for $\tan\beta=10$ and $\tilde m=300$
GeV,
\beq\label{eq:upmnp}
m_N\lsim1.4\times10^{12}\ {\rm GeV}.
\eeq
This bound scales like $(\sin2\beta/0.2)$. In the parameter region of
interest, the bound scales roughly as $(\tilde m/300\ {\rm
  GeV})^{2.17}$. (For related work on seesaw parameters in the PMSB
framework, see
\cite{Davidson:2001zk,Deppisch:2002vz,Masiero:2002jn,Ibarra:2009bg,Davidson:2011cb}.) 

\mysection{Consequences of observing $\mu\to e\gamma$}
If $\mu\to e\gamma$ is actually observed by MEG, this will have far
reaching consequences for GMSB. Explicitly, if BR($\mu\to e\gamma$) is
established to be within the range of Eq. (\ref{eq:meg}), we will be
able to make the following statements:
\begin{itemize}
\item The following lower bounds apply:
\beqa\label{eq:lomm}
M _m&\gsim&2\times10^{13}\ {\rm GeV\ or}\no\\
M_m&>&m_N\gsim3\times10^{11}\ {\rm GeV}.
\eeqa
Thus, low scale gauge mediation will be excluded.
\item At the same time, a bound similar to (and perhaps somewhat
  stronger than) Eq. (\ref{eq:upmmgm}) applies. Thus, the range of $M_m$
  will be determined to within about an order of magnitude.
\item Combining these considerations leads us to suspect that $x_{\mu
    e}$ of Eq. (\ref{eq:grga}) cannot be negligibly small. For
  example, taking $M_m\sim5\times10^{13}$ GeV and $
  m_N\sim10^{12}$ GeV, gives $x_{\mu e}\sim1$.
\end{itemize}

Thus, in case that $\mu\to e\gamma$ is observed and interpreted within
the GMSB framework, we should not neglect PMSB contributions to
$\delta_{\mu e}$. In particular, in Eq. (\ref{eq:brmeg}), we should
replace
\beq\label{eq:meggagr}
\delta_{\mu e}^2\to (\delta_{\mu e}+\delta_{\mu e}^{L,{\rm
    grav}})^2+(1/16)(\alpha_1/\alpha_2)^2(\delta_{\mu e}^{R,{\rm
    grav}})^2,
\eeq
where $\delta_{\mu e}$ stands for the pure gauge-mediation
contribution of Eq. (\ref{eq:mlga}), while $\delta_{\mu e}^{L,{\rm
    grav}}$ and $\delta_{\mu e}^{R,{\rm grav}}$ stand for the
contributions from gravity mediation to, respectively,
$(m^2_{\widetilde L})_{21}$ and $(m^2_{\widetilde E})_{21}$, both of
which we take to be estimated by Eq. (\ref{eq:mgrav}). Within our
framework, the contribution from $\delta_{\mu e}^{R,{\rm grav}}$ is
thus of order one percent of that from $\delta_{\mu e}^{L,{\rm
    grav}}$, and can be safely neglected for most purposes.

\begin{figure}[hbp]
\includegraphics[width=0.45\textwidth]{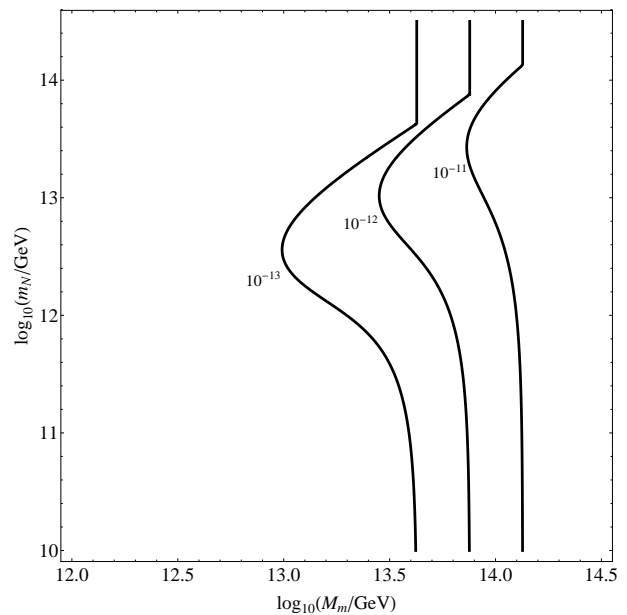}
\caption{Curves of fixed BR$(\mu\to
  e\gamma)=10^{-11},10^{-12},10^{-13}$ in the $M_m-m_N$ plane.}
\label{fig:mmmn}
\end{figure}

The interplay between gauge-mediated and gravity-mediated
contributions can be understood from Fig. \ref{fig:mmmn}. (A similar
figure, which neglects, however, the PMSB contributions, was presented
in Ref.  \cite{Tobe:2003nx}.) We can make the following statements
regarding an explanation of BR$(\mu\to e\gamma)>10^{-13}$ within our
framework:
\begin{itemize}
\item When $m_N> M_m$, the gravity-mediated contributions dominate
  $\mu\to e\gamma$. They can be large enough if
  $M_m\gsim2\times10^{13}$ GeV.
\item When $m_N\lsim3\times10^{11}$ GeV, gauge-mediated contributions are too
  small.  The decay rate can still be explained by gravity-mediated
  contributions if $M_m\gsim2\times10^{13}$ GeV.
\item For $M_m\gsim5\times 10^{13}$ GeV, gravity mediated contributions are,
  in general, too large.
\item For $10^{11}\ {\rm GeV}\lsim m_N < M_m \lsim10^{14}$ GeV,
  gauge-mediated contributions can be large enough.
\end{itemize}

Given this situation, we can now make a more precise statement about
$x_{\mu e}$. For a given value of BR$(\mu\to e\gamma)$, the $x_{ij}$
parameters are minimized when $\ln(M_m/m_N)=3/4$ and $M_m$ takes the
minimal value consistent with the decay rate, which we denote by $\hat
M_m$. From Eq. (\ref{eq:brmegmm}) we obtain (neglecting the
uncertainty in $\mu_{\mu e}$)
\beqa\label{eq:hatmm}
\frac{\hat M_m}{10^{13}\ {\rm GeV}}\simeq\sqrt{\frac{
{\rm BR}(\mu\to e\gamma)}{8\times10^{-13}}}\frac{20}{\tan\beta}
\frac{\hat{m}_{\tilde{L}}^2}{(300\ {\rm GeV})^2}.
\eeqa
Putting this value in Eq. (\ref{eq:grga}), we find
\beqa\label{eq:xijmin}
x_{\mu e}^{\rm min}\simeq0.07\sqrt{\frac{
{\rm BR}(\mu\to e\gamma)}{8\times10^{-13}}}\frac{20}{n_5\tan\beta}
\frac{\hat{m}_{\tilde{L}}^2}{(300\ {\rm GeV})^2}.
\eeqa
We remind the reader that this minimal value of $x_{\mu e}^{\rm min}$
is obtained under the assumption that the gravity mediated
contribution to the slepton masses-squared is of order $(F_s/M_{\rm
  Pl})^2$. It can be further suppressed if this scale is accompanied
with small or hierarchical dimensionless coefficients.

In addition to setting a lower bound on $x_{\mu e}$, a measurement of
$\mu\rightarrow e\gamma$ yields both upper and lower bounds on $M_m$,
as can be seen in Fig. \ref{fig:mmmn}. The upper bounds for the cases
of $n_5=1,20$ are given in Eq. (\ref{eqn:upperMmn5eq1}).
To determine the lower bound we require $\frac{\partial}{\partial
  m_N}\delta_{\mu e}=0$. Along a curve of constant $Br(\mu\rightarrow
e\gamma)$ this occurs at $\ln\frac{M_m}{m_N}=1$ (for which $x_{\mu e}$
is larger than the bound given by Eq. \ref{eq:xijmin}). The resulting
lower bound is:
\beq
\frac{M_m}{M_{Pl}}\gtrsim\begin{cases}
  3\times10^{-6}&n_5=1,\\
  2\times10^{-4}&n_5=20,\end{cases}
\eeq
where the bounds scale like $\sqrt{\frac{Br(\mu\rightarrow
    e\gamma)}{2.4\times
    10^{-12}}}\left(\frac{\hat{m}_{\tilde{L}}}{300\;{\rm GeV}}\right)^2
\left(\frac{20}{t_{\beta}}\right)$.
We conclude that a measurement of $Br(\mu\rightarrow e\gamma)$ will
fix $M_m$ to within one order of magnitude.

\mysection{Related observables}
The GMSB framework, combined with the MLFV seesaw mechanism, relate the
$\mu\to e\gamma$ decay rate to a number of other low energy
observables. In this section we review and update these relations in
view of the new measurements.

{\bf 1. $(g-2)$ of the muon:}\\
The dependence of ${\rm BR}(\mu\to e\gamma)$ on the supersymmetric
flavor-conserving parameters $\tan\beta$ and $m_{\widetilde L}$, Eq.
(\ref{eq:brmeg}), can be eliminated by the use of the
supersymmetric contribution to $(g-2)$ of the muon
\cite{Hisano:2001qz}:
\beq\label{eqgtwo}
{\rm BR}(\mu\to e\gamma)=3\times10^{-5}\left(\frac{\delta a_\mu^{\rm
      SUSY}}{10^{-9}}\right)^2 \delta_{\mu e}^2.
\eeq
Future experimental and theoretical developments will decide whether
it is more useful to use collider measurements of $\tan\beta$ and
$m_{\widetilde L}$, or the measurement of $(g-2)$, to extract
$\delta_{\mu e}$ from ${\rm BR}(\mu\to e\gamma)$ and by that test the
GMSB framework.

{\bf 2. $\mu\to e$ conversion:}\\
An interesting observable is the $\mu-e$ conversion rate $R(\mu\to
e)$. In the GMSB framework, the photon penguin diagram dominates both
the radiative decay and the $\mu\to e$ conversion, and thus the two
rates are related in a way that is independent of the model
parameters:
\beq
\frac{R(\mu\to e\ {\rm in\ Ti})}{{\rm BR}(\mu\to e\gamma)}\simeq0.005.
  \eeq
Clearly, all the implications that we describe here for a signal in
the $\mu\to e\gamma$ decay can be made in a similar way from a signal
of $\mu\to e$ conversion.

{\bf 3. Radiative $\tau$ decays:}\\
Models with quasi-degenerate sleptons correlate all three radiative
charged lepton decay rates in a simple way:
\beq\label{eq:radrat}
{\rm BR}(\tau\to \ell\gamma)=0.175{\rm BR}(\mu\to
e\gamma)(\delta_{\tau\ell}/\delta_{\mu e})^2.
\eeq
GMSB in its version employed here translates Eq. (\ref{eq:radrat})
into an expression that depends only on light neutrino flavor
parameters:
\beq\label{eq:radratmu}
{\rm BR}(\tau\to \ell\gamma)=0.175{\rm BR}(\mu\to
e\gamma)(\mu_{\tau\ell}/\mu_{\mu e})^2.
\eeq
For our two sets of mixing parameters, we get the following
predictions:
\beqa\label{eq:brbr}
\frac{{\rm BR}(\tau\to e\gamma)}{0.175{\rm BR}(\mu\to
e\gamma)}&\simeq&0.04-0.17,\no\\
\frac{{\rm BR}(\tau\to \mu\gamma)}{0.175{\rm BR}(\mu\to
e\gamma)}&\simeq&5.4-9.0.
\eeqa

Whether the gravity mediated contributions are significant depends on
the $x_{ij}$ parameters. Here, our framework gives
\beq\label{eq:ratxij}
\frac{x_{\tau\ell}}{x_{\mu e}}\sim\frac{\mu_{\mu e}}{\mu_{\tau\ell}},
\eeq
which leads to
\beqa\label{eq:xtau}
{x_{\tau e}}/{x_{\mu e}}&\simeq&2.2-4.0,\no\\
{x_{\tau\mu}}/{x_{\mu e}}&\simeq&0.33-0.42.
\eeqa
We learn that our framework predicts ${\rm BR}(\tau\to\mu\gamma)
\gsim{\rm BR}(\tau\to e\gamma)$ and, furthermore, among the three rates of
radiative decays, $\tau\to\mu\gamma$ is the closest to represent the
seesaw flavor parameters.

\mysection{Conclusions}
Within the framework of gauge mediated supersymmetry breaking, flavor
violation is minimal. Yet, observable lepton flavor violation can be
generated if the seesaw scale is lower than the scale of gauge
mediation. In particular, a measurement of the $\mu\to e\gamma$ decay
rate is sensitive to the gauge-mediation and seesaw scales, and to the
seesaw parameters.

In this work, we focus our attention on ``ordinary gauge mediation'',
with weakly coupled messenger fields, and on ``minimally lepton flavor
violating'' seesaw sector, with degenerate singlet neutrinos and real
Yukawa couplings. Many of our conclusions (such as the significance of
Planck scale mediation) hold in more generality, but we postpone a
detailed study of these generalizations to future work.

The recent MEG upper bound, BR($\mu\to e\gamma)\leq2.4\times
10^{-12}$, gives the strongest upper bound up-to-date on the mediation
scale, which for a small number of messenger fields reads
$M_m\lsim10^{14}$ GeV [see Eq.  (\ref{eqn:upperMmn5eq1})]. The seesaw
scale is either higher than the gauge mediation scale or lower than
about $10^{13}$ GeV [see Eq.  (\ref{eq:upimp})]. The bound on the
seesaw scale in the framework of gravity mediated supersymmetry
breaking is of order $10^{12}$ GeV.

If the $\mu\to e\gamma$ decay is eventually observed by MEG, it will
determine the scale of gauge mediation to within one order of
magnitude. It will be impossible, however, to interpret the
measurement in terms of purely seesaw parameters, because
contributions from gravity mediation will be, in general, significant.

\mysections{Acknowledgments} We thank Yonit Hochberg for useful
discussions. YN is the Amos de-Shalit chair of
theoretical physics and supported by the Israel Science Foundation,
and the German-Israeli foundation for scientific research and
development (GIF).


\end{document}